\documentclass[aps,prb,preprint,groupedaddress,showpacs,amsmath]{revtex4}

\usepackage{amsfonts}
\usepackage{graphicx}
\begin{document}

\title{Experimental investigation of flux motion in exponentially shaped
Josephson junctions}

\author{G. Carapella}
\thanks{Corresponding author}
\email[\newline e-mail: ]{giocar@sa.infn.it}
\thanks{\newline FAX: +3908965390 \\}
\affiliation{Departement of Physics \lq\lq E. R. Caianiello\rq\rq
, and INFM Research Unit, University of Salerno, via
S. Allende, I-84081 Baronissi, Italy. }
\author{N. Martucciello}
\affiliation{Departement of Physics \lq\lq E. R. Caianiello\rq\rq
, and INFM Research Unit, University of Salerno, via
S. Allende, I-84081 Baronissi, Italy. }

\author{G. Costabile}
\affiliation{Departement of Physics \lq\lq E. R. Caianiello\rq\rq
, and INFM Research Unit, University of Salerno, via
S. Allende, I-84081 Baronissi, Italy. }

\date{\today}

\begin{abstract}
We report experimental and numerical analysis of
expontentially shaped long Josephson
junctions with lateral current injection. Quasi-linear
flux flow branches are observed in the current-voltage characteristic of
the junctions in the absence of magnetic field. A
strongly asymmetric response to an applied magnetic field is also
exhibited by the junctions.
Experimental data
are found in agreement with numerical predictions
 and demonstrate
the existence
of a geometry-induced potential  experienced
by the flux quanta in nonuniform width junctions.

\end{abstract}


\maketitle

\section{Introduction}

In recent years, there have been some theoretical studies concerning the
possibility to influence the flux motion in long Josephson junctions by
mean of geometry-induced or field-induced potentials.
The most known example is the annular
junction embedded in a spatially homogeneous magnetic field. As it is known
\cite{Nie2}, in this case a cosinusoidal potential is experienced by a flux
quantum trapped in the junction when a spatially homogeneous magnetic field
is applied parallel to the junction barrier. The origin of the potential is
caused by the spatial variation of the radial component of the magnetic
field that in this special geometry becomes sinusoidal.
A field-induced sawtooth-like potential has been recently
considered \cite{Cara} for experimental demonstration
\cite{Cara1} of ratchet effect
in annular junctions. Currently, modifications
of the annular geometry, as the heart-shaped geometry
\cite{Ustin} are used to achieve a field-induced
double-well potential for demonstration of fluxon quantum-bit and
macroscopic quantum coherence phenomena.

To achieve potentials without the help of a magnetic field, the case of a
nonuniform junction width has been theoretically addressed in recent years
for linear \cite{Sakai,Sergio,Scott,Ben} as well as for annular
\cite{Gol} geometries. Theoretically,
a geometry-induced potential related to the
spatial variation of the junction width is expected. This potential
corresponds to a force acting on the fluxons in the direction of the
shrinking width. Recently, \cite{Koshe} ordinary Josephson flux-flow
oscillators \cite{Giap,Koshelets} have been modified adding to the classical
overlap geometry \cite{Barone} unbiased pointed tails. This is expected to
enhance the annichilation of the fluxons at the edges of the oscillator,
with consequent reduction of fine structures in the ordinary
velocity-matching step \cite{Giap,Koshelets,Cirillo}.

In this paper we experimentally address the existence of the geometrical
force in nonuniform width junctions.  To do this, we consider an
exponentially shaped overlap junction with lateral current injection. The
lateral current injection acts as a flux quanta generator, also in the
absence of a magnetic field, while the unbiased shaped region should act as
an accelerating region for both fluxons or antifluxons injected at one edge
of the junction. If really present, the geometric force should allow to
achieve a quite regular unidirectional flux flow motion in the junction,
without the help of a magnetic field. This dynamical regime should be
accounted for a  branch in the current-voltage curve of the
shaped junction. The demonstration of such branches in the current-voltage
curves of shaped junctions we fabricated suggests that such a geometrical
force is really experencied by the flux quanta.

The paper is organized as follows. In Sec. II we specialize the general
model \cite{Sergio,Scott} for a long junction with nonuniform width to our
exponentially shaped junction with lateral current injection. In Sec. III
the experimental results both in the absence of magnetic field and in the
presence of magnetic field are presented and discussed with the help of
numerical simulations. Finally, main results are summarized in the
Conclusions.

\section{Theory}

\begin{figure}
\includegraphics[width=8cm,clip=]{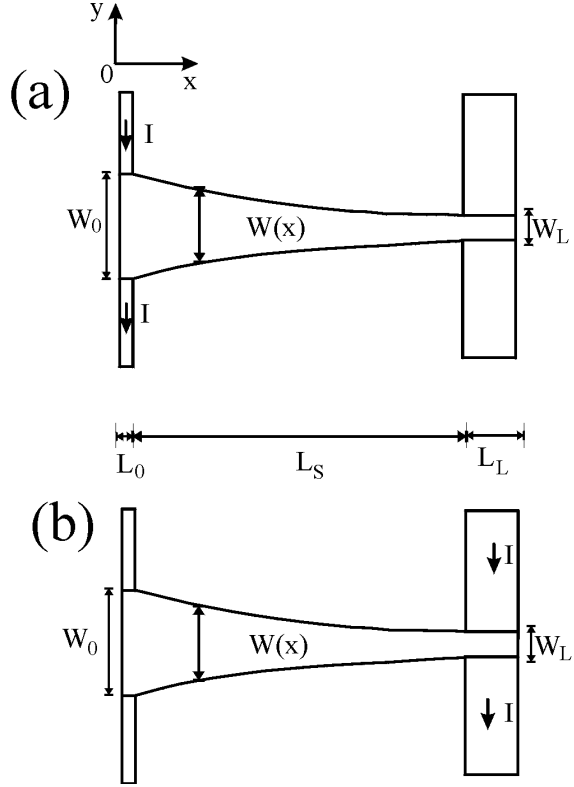}
 \caption{Exponentially shaped junction with lateral current
 injection. The current can be fed into the left [(a)] or into the
 right [(b)] edge.}
 \label{fig1}
 \end{figure}

Under some simplificative hypotheses \cite{Sergio}, the model for a long overlap
junction with nonuniform width $W(x)=f_{1}(x)-f_{2}(x)$ was found as
\cite{Sergio,Scott}

\begin{equation}
\phi _{xx}-\phi _{tt}=\sin \phi +\alpha \phi _{t}-\frac{W^{\prime }(x)}{%
W(x)}\phi _{x}+\eta _{y}\frac{W^{\prime }(x)}{W(x)}-\Gamma (x),  \label{uno}
\end{equation}
with 
\begin{equation}
\Gamma (x)=\frac{\left. \eta _{x}\right| _{f_{2}}-\left. \eta _{x}\right|
_{f_{1}}}{W(x)}.  \label{due}
\end{equation}
In Eq.~(\ref{uno}) $\phi $ is the Josephson phase, $\alpha $ is the
dissipation parameter, $\eta _{x}$ and $\eta _{y}$ are the normalized
magnetic fields in the $x$ and $y$ directions, respectively. Space is
normalized to the Josephson penetration length $\lambda _{J}$ and time to
the inverse of the plasma frequency $\omega _{J}=\overline{c}/\lambda _{J},$
with $\overline{c}$ the velocity of electromagnetic waves in the junction.
For the geometry we report here (see Fig.~\ref{fig1}) the total physical
length of the junction is $L=L_{0}+L_{S}+L_{L}\gg \lambda _{J}$, while the
width is chosen as
\begin{equation}
W(x)=\left\{ 
\begin{array}{l}
W_{0}\qquad 0<x\leq L_{0}, \\ 
W_{0}\exp \left[ \frac{1}{L_{S}}\ln (\frac{W_{L}}{W_{0}})(x-L_{0})\right]
\qquad L_{0}<x\leq L_{S}+L_{0,} \\ 
W_{L}\qquad L_{S}+L_{0}<x\leq L,
\end{array}
\right.   \label{unobis}
\end{equation}
The bias current $I$ can be fed into the left or into the right edge, as
shown in Fig.~\ref{fig1}.
The bias term $\Gamma (x)$ in Eq. (\ref{due}) becomes

\begin{equation}
\gamma _{A}(x)=\left\{ 
\begin{array}{l}
\frac{IL}{J_{0}L_{0}W_{0}L}\equiv \gamma \frac{L}{L_{0}}\qquad 0<x\leq L_{0,}
\\ 
0\qquad L_{0}<x\leq L,
\end{array}
\right.   \label{tre}
\end{equation}
for left edge current injection, and 
\begin{equation}
\gamma _{B}(x)=\left\{ 
\begin{array}{l}
0\qquad 0<x\leq L_{S}+L_{0}, \\ 
\frac{IL}{J_{0}L_{L}W_{L}L}\equiv \gamma \frac{L}{L_{L}}\qquad
L_{S}+L_{0}<x<L,
\end{array}
\right.   \label{quattro}
\end{equation}
for right edge current injection.
The lengths $L_{0},$ $L_{L},W_{0},$and $W_{L}$ are chosen to be shorter than 
$\lambda _{J}$ and such that $L_{L}W_{L}=L_{0}W_{0},$ to have equal biased
areas.

Hence, the model for our exponentially shaped junction becomes
\begin{subequations}
 \label{model}
\begin{eqnarray}
\phi _{xx}-\phi _{tt} &=&\sin \phi +\alpha \phi _{t}+\lambda \phi _{x}-\eta
\lambda -\gamma _{A,B}(x),   \\
\varphi _{x}(0) &=&\eta ,   \\
\varphi _{x}(l) &=&\eta ,  
\end{eqnarray}
\end{subequations}
where $l=L/\lambda _{J},$ $\lambda =\lambda _{J}\ln (W_{0}/W_{L})/L_{S}$,
and $\eta $ accounts for an external magnetic field applied in the $y$
direction. We remark that the chosen geometry  reduces to the
exponentially shaped in-line geometry \cite{Scott,Ben} in the case of left
current injection with injection length $L_{0}$ much lower than $%
\lambda _{J}.$ In such a limit, the model becomes
\begin{subequations}
 \label{modelin}
\begin{eqnarray}
\phi _{xx}-\phi _{tt} &=&\sin \phi +\alpha \phi _{t}+\lambda \phi _{x}-\eta
\lambda ,   \\
\varphi _{x}(0) &=&\eta -\gamma l,   \\
\varphi _{x}(l) &=&\eta .  
\end{eqnarray}
\end{subequations}
As first noted in Ref. \onlinecite{Sergio},
a  force that drags fluxons or antifluxons in
the direction of the narrowing width is expected for our geometry. In fact,
in the absence of magnetic field ($\eta =0$), for a soliton

\[
\phi (x)=4\arctan \left[ \exp \left[ \sigma \frac{x-ut}{\sqrt{1-u^{2}}}%
\right] \right] 
\]
in the unbiased region [$\gamma _{A,B}(x)=0]$, the following equation of
motion can be found from Eqs.~(\ref{model}) in the framework of the
perturbative approach \cite{Scottve}

\begin{equation}
(1-u^{2})^{-3/2}\frac{du}{dt}=-\alpha \frac{u}{\sqrt{1-u^{2}}}+\frac{\lambda 
}{\sqrt{1-u^{2}}}.  \label{motion}
\end{equation}
This indicates that, if present, both a fluxon or an antifluxon will
experience a  force proportional to the shaping parameter $\lambda $ and
will be accelerated in the narrowing width direction, i.e. from the left to
the right in our geometry. From Eq. (\ref{motion}) the stationary velocity
of the motion will be $u=\lambda /\alpha $ for $\lambda /\alpha <1$ or the
limit velocity $u=1$ for $\lambda /\alpha >1.$

\section{Numerical and experimental results}

We realized Nb/Al$_{2}$O$_{3}$/Nb junctions with the geometry shown in
Fig.~\ref{fig1}. The physical dimensions of the junctions were $%
L=L_{0}+Ls+L_{L}=(10+560+40)$ $\mu $m, $W_{0}=40$ $\mu $m, $W_{L}=10$ $\mu $%
m. For the two junctions we report here the normalized lengths were $%
l\approx 20,$ and $l\approx 17,$ with shaping parameters $\lambda \approx
0.07$ and $\lambda \approx 0.08$, respectively. In the following the
behavior of the junctions in the absence of magnetic field as well as the
response to a magnetic field applied along the $y$ direction is discussed.

\subsection{Behavior in the absence of magnetic field}

\begin{figure}
\includegraphics[width=8cm,clip]{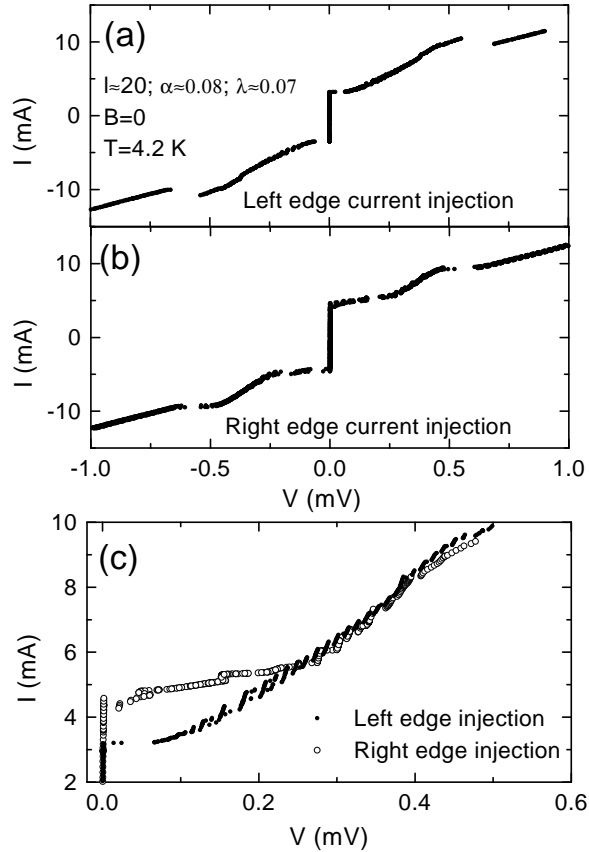}
 \caption{Current-voltage curve of a junction with left current
 injection (a) or with right current injection (b) exhibiting
 quasi-linear flux flow branches. In (c) the flux
 flow branch for left current injection is compared with the
 one achieved for right current injection.}
 \label{fig2}
 \end{figure}

The current-voltage curve for the junction with $l\approx 20$ is reported in
Fig.~\ref{fig2}. No external magnetic field is applied to the junction.
Panel (a) of
this figure refers to the case of current injected into the left edge, while
the panel (b) refers to the case of current injected into the right edge. In
both cases an almost linear branch starting from a
certain critical current
and extending for a given current range is observed, but some qualitative
differences exist between the two cases. As better seen in Fig.~\ref{fig2}(c), when
current is injected into the left edge the branch starts to a lower critical
current and is more regular than the branch achieved with current injected
to the right edge. The observed behavior seems to be consistent with the
idea that a geometrical force is effectively experienced by the fluxons in
this geometry. When current of positive (negative) polarity is injected into
the left edge, antifluxons (fluxons) will enucleate at this edge after a
critical current value has been reached. From Eq. (\ref{motion}) these
antifluxons are expected to be accelerated toward the other edge because of
the geometrical force, so realizing a regular unidirectional flux motion.
Conversely, if the current is injected into the right edge, the antifluxons
enucleated a this edge should overcome a force to travel toward the left
edge. This should result in an irregular or chaotic flux motion. Moreover,
in the left current injection the geometrical force helps to start an
antifluxon (or fluxon) motion, corresponding to a voltage in the I-V curve
beyond a critical current value. In the right current injection such a force
 opposes the starting of the flux motion, so resulting in a larger
critical current.

We should remark that the quasi-linear branches reported in Fig.~\ref{fig2}
 could
remind the Displaced Linear Slope branches sometimes
reported for rectangular in-line or overlap
geometries \cite{Barone1,Scott1,Pace,Usthen,Cirillo1}.
However here the branches are
rather regular, quite noiseless, and are obtained in the
absence of magnetic field.

\begin{figure}
\includegraphics[width=8cm,clip]{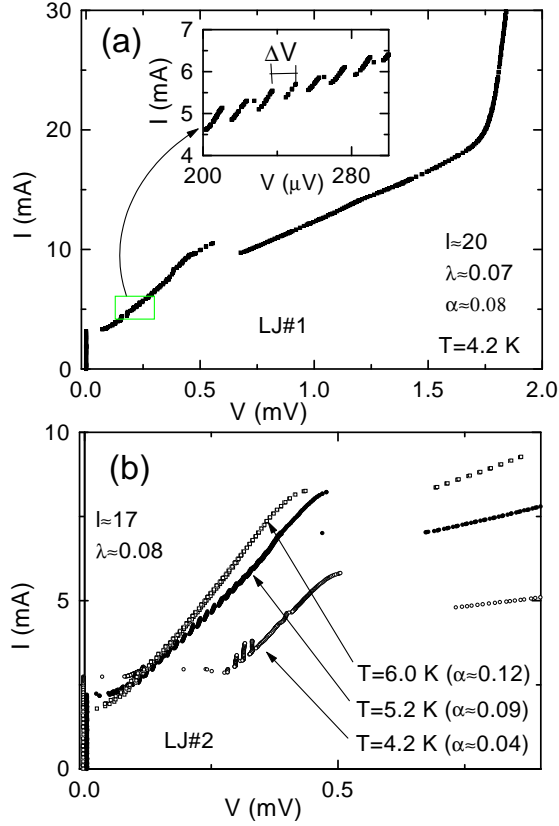}
 \caption{(a) The current-voltage curve of Fig.~\ref{fig2}(a) is replotted
 on a larger scale. Small steps in the flux-flow branch are shown
 in the inset. (b) Flox flow branch of a junction with $\lambda \approx 0.08$
 at different temperatures.}
 \label{fig3}
 \end{figure}

In the following we will focus on the left current injection.
In Fig.~\ref{fig3}(a)
same data of Fig.~\ref{fig2}(a) are replotted on a larger scale.
As better seen in
the inset, the flux flow branch exhibits a series of small steps spaced of a
voltage $\Delta V\approx 15$ $\mu $V. The curve reminds the numerically
predicted curve for an exponential shaped asymmetric in-line junction
\cite{Ben}.
The observed voltage spacing is consistent with the spacing of cavity mode
resonances expected from physical dimensions of our junction, $\Delta V$ $%
\simeq \Phi _{0}\overline{c}/L.$ This is to be expected due to the open
boundary condition at the edges. The antifluxons in the chain moving toward
the right edge will be reflected as fluxons. If the dissipation $\alpha $ is
not too large, the antifluxons can have enough energy to travel toward the
left edge as fluxons after the reflection, despite of the geometrical force
opposing the motion. This mechanism can excite cavity mode resonances.
However, for quite large dissipation one expect that the reflected motion
would be more and more damped, and the excitation of cavity modes should be
consequently damped. In Fig.~\ref{fig3}(b) the I-V curve of the junction with $%
l\approx 17$ is plotted for three different temperatures, corresponding to
three different $\alpha $ values. As  expected, the small steps accounting
for the cavity mode resonances are more and more damped as the dissipation
(temperature) is increased.

\begin{figure}
\includegraphics[width=8cm,clip]{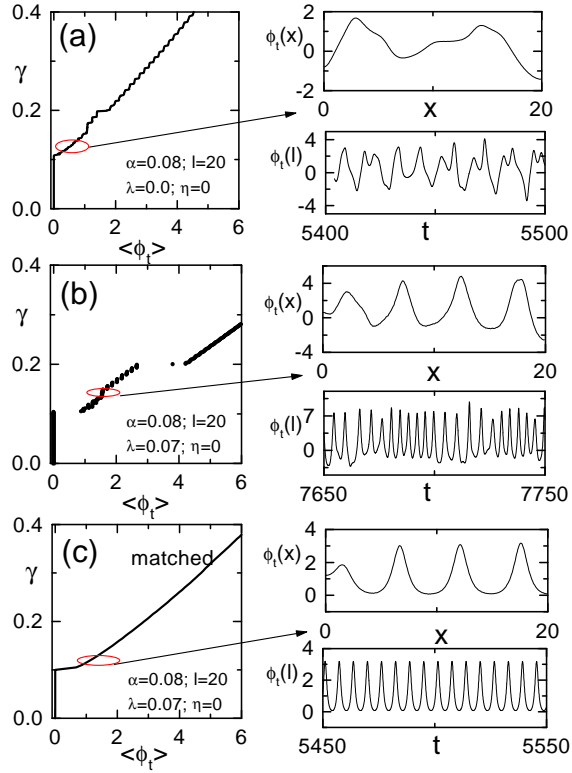}
 \caption{Calculated current-voltage curve for an
 unmatched rectangular [(a)],
 for an unmatched exponentially shaped [(b)], and for a matched exponentially
  shaped [(c)] junction with lateral current injection.
 The snapshots shows the instantaneous voltage profiles and the
 voltage signal at the right edge of the junctions.}
 \label{fig4}
 \end{figure}

To gain further insight in the flux motion dynamics in the absence of
magnetic field, we integrated the model Eqs.~(\ref{model}) with $\eta =0$ and
forcing term $\gamma _{A}(x)$ defined in Eq.~(\ref{tre}). In
Fig.~\ref{fig4}(a) we
show the calculated current voltage curve for a junction with left edge
current injection and of uniform width [$\lambda =0$ in Eqs.~(\ref{model})].
This is equivalent to the asymmetric in-line rectangular
geometry \cite{Barone}.
As seen in the snapshot showing the instantaneous
 voltage distribution in
the junction, antifluxons are created at the biased edge. However, due to
the absence of a force in the unbiased region, the flux motion is not very
regular. Here it is only the repulsion between flux quanta that tends to
drive the flux toward the right edge. The resulting motion is quite noisy,
as well as the calculated ac voltage at the right edge.

In Fig.~\ref{fig4}(b) the case of an
exponentially shaped width ($\lambda =0.07$) is
considered. In the simulation we used parameters similar to the ones
estimated for the experimental curve in Fig.~\ref{fig3}(b). As it seen in the
snapshot, now the presence of a force in the unbiased region makes the flux
motion toward the right edge more regular, as well as the voltage signal at
the right edge. In the snapshot four moving antifluxon are counted as
present in the mean in the junction for the chosen bias point. By increasing
the bias current, more and more antifluxons can be injected in the junction,
and the chain becomes more and more dense. Correspondly, the voltage signal
at the right edge becomes less impulsive and approaches a sinusoidal form.
When the junction is completelly filled with antifluxons, a transition to a
new dynamical regime, similar to a laminar phase flow \cite{Ben}
is achieved. In
the current-voltage curve, this transition corresponds to the switch from
the  flux flow branch toward the other resistive branch, beyond
a critical current value. As in the experimental curve, the small steps in
the calculated flux flow branch are spaced of $\pi /l$, the spacing (in
normalized units) expected from cavity mode resonances excited from
reflecting boundary conditions at the edges. 

As said above, the resonances can be damped by increasing the dissipation $%
\alpha .$ However, another mean to achieve this is to match the impedance of
the fluxon chain with a load $z$ at the right edge. The matching is found
possible \cite{Scott,Ben} for this geometry, conversely to the rectangular
asymmetric in-line geometry. For our exponentially shaped junction the wave
impedance is just the stationary velocity we have found above, $-\phi
_{t}/\phi _{x}=u=\lambda /\alpha .$ This should be matched with a load of
impedance $z=-\phi _{t}(l)/\phi _{x}(l).$ Numerical results for the case of
a matched load are shown in Fig.~\ref{fig4}(c). As a result of the absence of
reflections, the small steps typical of the unmatched case of
Fig.~\ref{fig4}(b) are
now absent and the flux chain exhibits a very
regular motion, as well as the
voltage signal at the matched edge. This characteristic makes the
exponentially shaped junction interesting as zero-magnetic field flux flow
oscillator.

\begin{figure}
\includegraphics[width=8cm,clip]{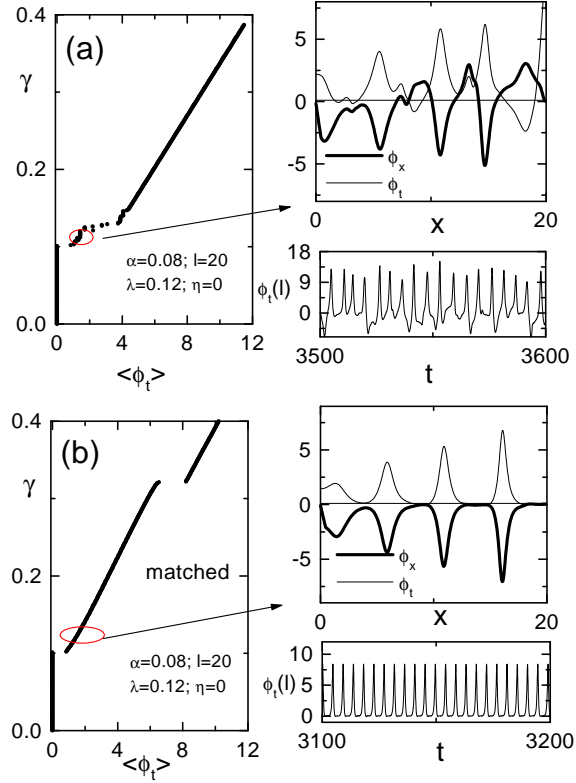}
 \caption{Calculated current-voltage curve for 
  an unmatched [(a)] and a matched [(b)] exponentially
  shaped junction with $\lambda =0.12$.
 The snapshots shows the instantaneous voltage and
 magnetic field profiles, and the
 voltage signal at the right edge of the junctions.}
 \label{fig5}
 \end{figure}

In Fig.~\ref{fig5} there are shown numerical results
for a junction with $\lambda /\alpha
>1.$ For the unmatched case shown in Fig.~\ref{fig5}(a)
the used parameters can
account for the current-voltage curve at T=4.2 K in Fig.~\ref{fig3}(b).
In both cases
$\lambda /\alpha >1$. As said above, in such a case the antifluxons are
accelerated toward the asymptotic velocity $u=1,$ so the matching condition
is now $z=1.$ In the snapshots shown in Fig.~\ref{fig5}(a) and (b)
we show also the
instantaneous magnetic field. Antifluxons in the chain are accelerated and,
due to their relativistic nature, they are Lorentz-contracted as the
asymptotic velocity is approached. As for the case $\lambda /\alpha <1,$
the matched junction exhibits a smooth flux-flow branch and the voltage
signal at the right edge is very regular.

\subsection{Behavior in the presence of magnetic field}

\begin{figure}
\includegraphics[width=8cm,clip]{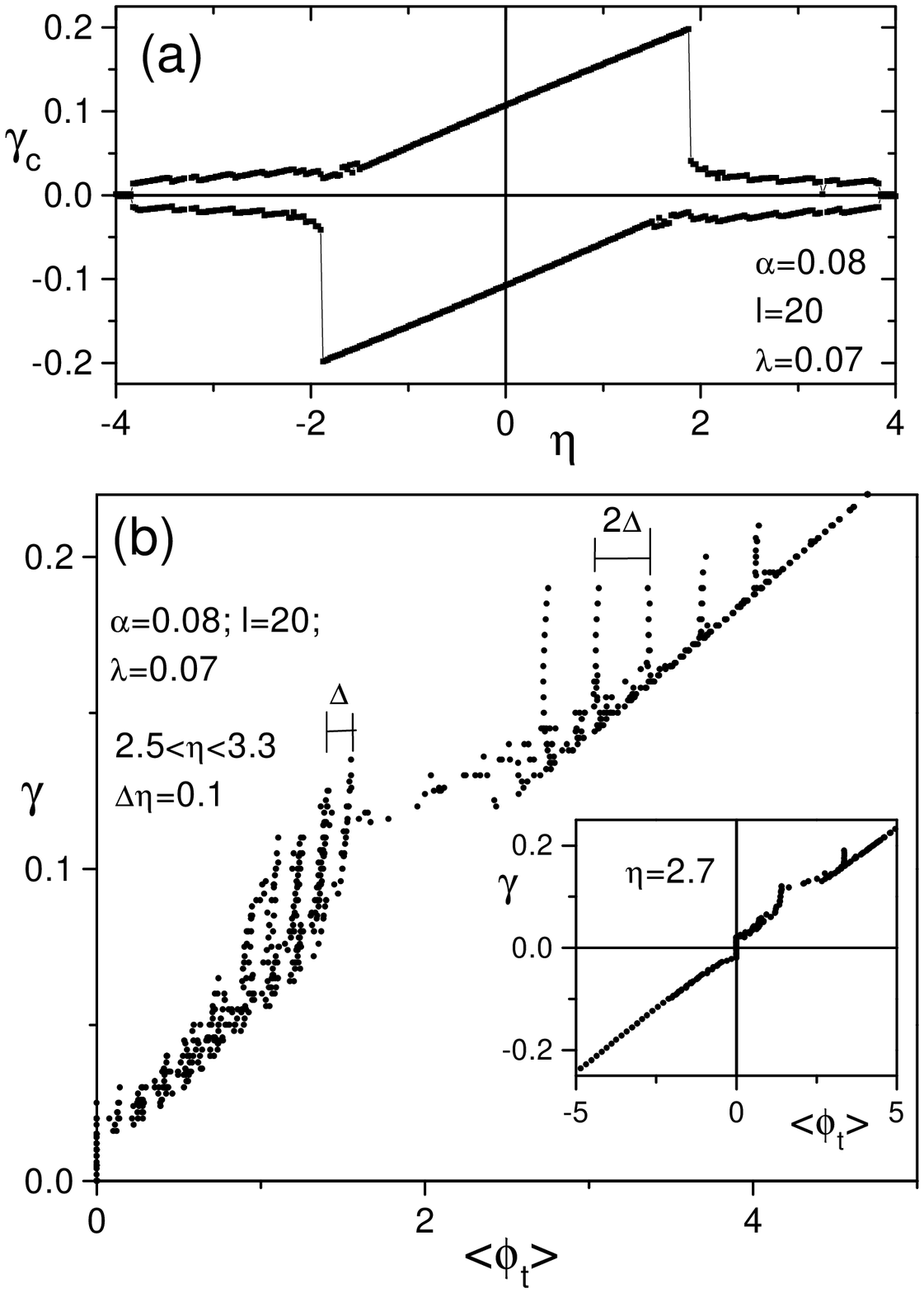}
 \caption{(a) Calculated magnetic field pattern for a junction
 with $\lambda =0.07$. (b) Calculated steps induced in
 the current voltage characteristic by a magnetic field varied
 in the range $2.5<\eta <3.3$.
 In the inset the steps at for a given field value are shown.}
 \label{fig6}
 \end{figure}

Due to the lateral current injection and due to the existence of a preferred
direction of motion, our exponentially shaped junction is expected to show a
behavior in magnetic field even more strongly asymmetric with respect to the
asymmetric rectangular in-line geometry. The calculated critical current as
a function of the magnetic field is shown in Fig.~\ref{fig6}(a) for
a junction with $%
l=20$ and $\lambda =0.07$. The pattern was obtained integrating
Eqs.~(\ref{model}) with $\eta \neq 0.$ A quite abrupt decrease of the critical current
around $\eta =2$ is recovered. In normalized units, for this value of
magnetic field a fluxon is enucleated in the junction. The almost complete
absence of secondary lobes in the pattern means the almost complete absence
of trapped flux in this kind of junction, an indication that the geometrical
force helps to move fluxons just after their enucleation. 

Figure \ref{fig6}(b) shows a global representation
of calculated resonant steps
achieved for normalized magnetic field values slightly larger than the
critical value $\eta =2.$ Such a plot is obtained superimposing the curves
corresponding to different values of magnetic field. Two families of steps
with two characteristic voltage spacing are recovered. The lower voltage
spacing is the one expected for Fiske mode resonances, $\Delta \sim \pi /l$,
the larger one is about $2\pi /l=2\Delta $. The family of steps with larger
voltage spacing appears in the same current range where the
\begin{figure}
\includegraphics[width=8cm,clip]{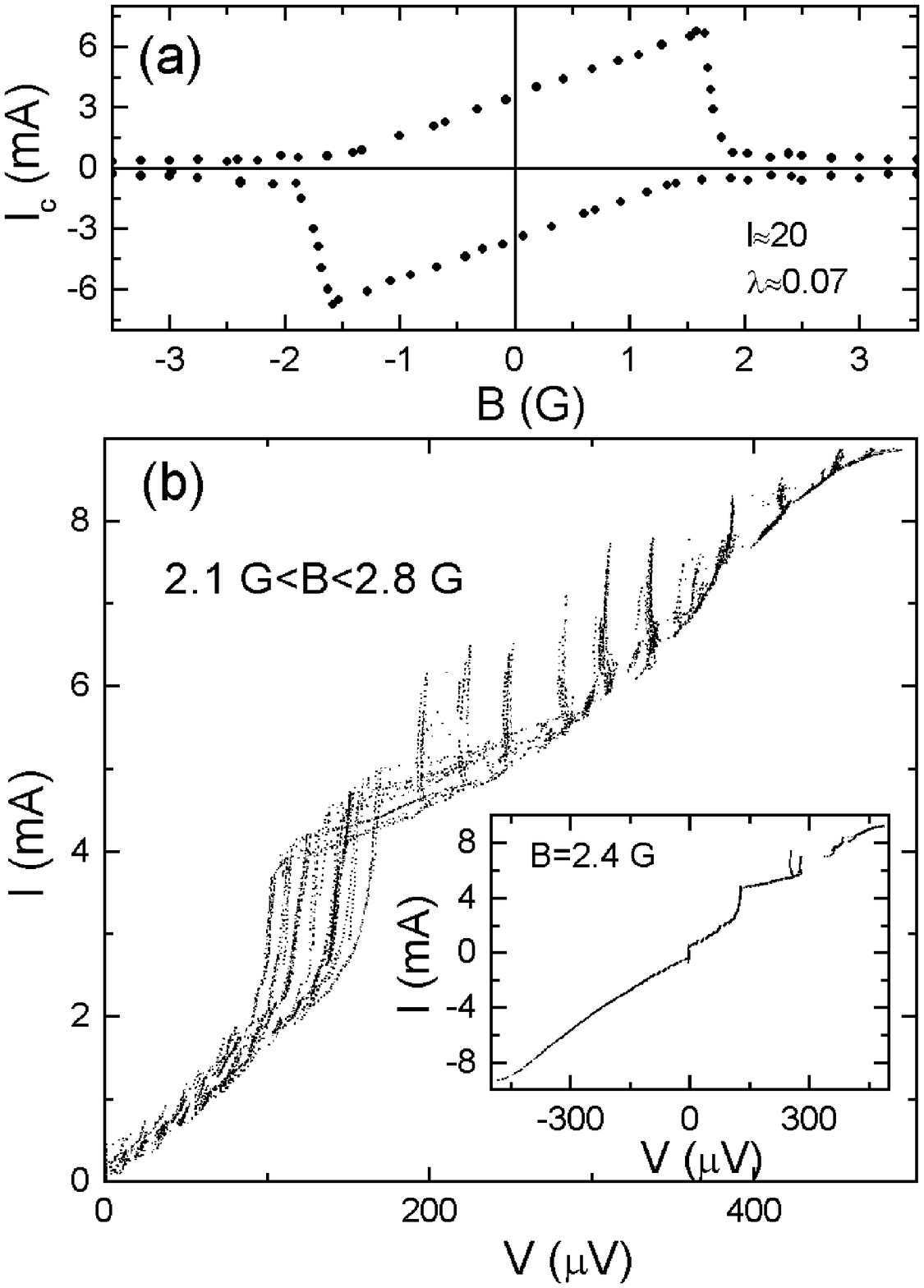}
 \caption{Measured magnetic field pattern (a) and
 current steps (b) induced in a junction
 with $\lambda \approx 0.07$ by a magnetic field slightly
 larger than the critical field. In the inset the current
 voltage curve at B=2.4~G is shown.}
 \label{fig7}
 \end{figure}
flux flow branch is achieved in the absence of magnetic field. From
numerical simulation it is seen that these steps with larger voltage spacing
consists of cavity mode resonances excited by a fluxon chain and an
antifluxon chain moving in opposite directions. In fact, for positive bias
current an antifluxon chain moving toward the rigth edge is generated, while
a positive magnetic field generate fluxons. These fluxons are pulled toward
the left edge by the Lorentz force associated to the positive bias current a
this edge. Antifluxons travelling toward the right generate a voltage with
positive polarity that adds to the voltage of the fluxons traveling toward
left. This accounts for the larger voltage spacing observed for the steps of
this family.

For negative bias currents, the fluxons generated by the positive magnetic
field are pushed toward the right edge by the Lorentz force associated to
the negative bias current. Moreover, also the fluxon chain injected at the
left edge by the bias current moves toward the right, under the action of
the geometrical force. The result is that an unidirectional motion toward
the right edge tends to be achieved, with consequent damping of resonant
steps accounting for cavity mode resonances. As seen in the inset of
Fig.~\ref{fig6}(b), resonant steps are in fact virtually absent
for negative current
values. This peculiarity is observed also for larger magnetic values.  For
negative magnetic fields values, the current-voltage curve show steps for
negative bias current values, and no steps for positive bias values.
In other words, the current-voltage curve reflects with respect to the
origin when the sign of the magnetic field is inverted.

\begin{figure}
\includegraphics[width=8cm,clip]{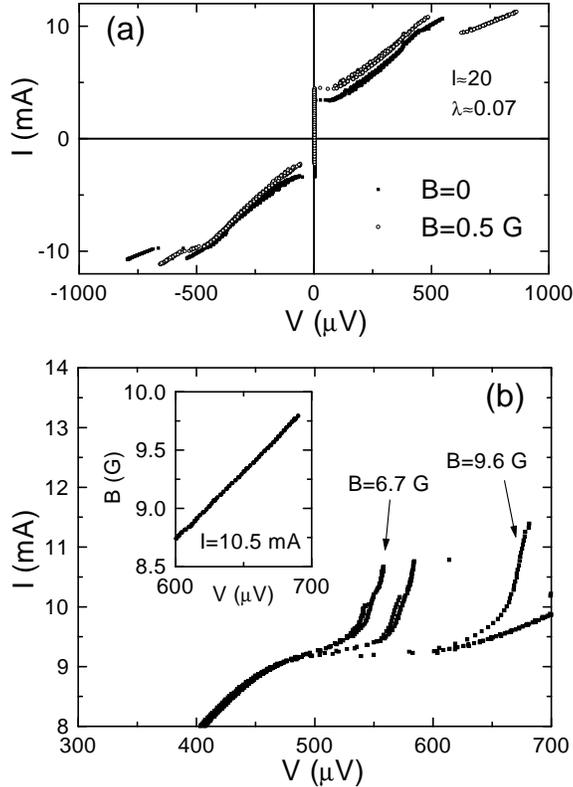}
 \caption{(a) Modification of the flux flow branch induced by a
 magnetic field lower than the critical field.
 (b) Steps recorded for magnetic fields quite larger than the critical field
 of the junction. A smooth single step
 is achieved for $ B> 8~G$. In the inset we show
 the voltage of the junction polarized
 at fixed current on the step as a function of the magnetic field.}
 \label{fig8}
 \end{figure}

Predicted strongly asymmetric magnetic field behavior summarized in
Fig.~\ref{fig6}
is fully recovered in the experiment, as shown in Fig.~\ref{fig7}.
Experimental
data refer to the junction with $l\approx 20$.
In Fig.~\ref{fig7}(b) the steps are achieved for magnetic field values
slightly larger than $B_{c}=1.75$ G, the critical field of the junction. In
Fig.~\ref{fig8}(a) we reported the modification of
the flux flow branch induced by a
low magnetic field, i.e., lower than $B_{c}.$ An asymmetric tunning of the
branch is recovered . This can be easily understood as follows. We are using
strongly left-edge-peaked current injection. In this case our junction can
be also described as a shaped aysmmetric in-line junction, i.e., with model
Eqs.~(\ref{modelin}). Looking at the vortex generator term $\phi _{x}(0)=\eta
-\gamma l,$ it is easily understood that a positive magnetic field
cooperates with a negative bias current to inject fluxons in the junctions,
while such a positive magnetic field opposes a positive current injecting
antifluxons in to the junction. The result is that, for a given absolute
value of the bias current, more solitons are present at negative polarities
than at positive polarities, resulting in a voltage at negative polarity
larger than the voltage at positive polarity, as it is in fact observed in
the experimental curves of Fig.~\ref{fig8}(a).

Figure \ref{fig8}(b) shows the steps recorded in the junction
for magnetic fields
quite larger than $B_{c}.$ The fine structures accounting for Fiske modes
are completely smeared out at $B\approx 9$ G and a smooth step, 
similar to the velocity-matching step \cite{Giap,Koshelets,Cirillo}, with
asymptotic voltage strictly proportional to $B$ is achieved. The
proportionality between the voltage and the magnetic field is shown in the
inset for a given biasing current. The velocity-matching step  is found to
exist beyond the current-voltage range of existence of the linear flux flow
branch in zero field, i.e., beyond $V=$600 $\mu $V (and up to 1500 $\mu $V)
and $I=9.5$ mA. This is consistent with the observation \cite{Cirillo}
that the
velocity matching step originates from a quasi-linear waves regime in the
junction. In our case such a quasi-linear background is the laminar phase
flow \cite{Ben} achieved beyond the current range typical of the zero-field
flux-flow branch.

\section{Conclusions}

Summarizing, we have experimentally investigated the occurence of dynamical
states in exponentially shaped overlap junctions with lateral current
injection. In zero magnetic field, the lateral current injection acts as a
fluxon or antifluxon chain generator and these chains can be accelerated by
the geometrical force originating from the nonuniform width of the junction.
The result is that a quite regular flux-flow motion, corresponding to a
quasi-linear branch in the I-V curve of the junction, can be achieved in
this kind of junction without the help of an external magnetic field.
Moreover, numerical simulations show that this kind of motion can be
precisely matched to a load, a peculiarity that makes the zero field flux
flow branch interesting for zero-field flux flow oscillators. In the
presence of a magnetic field a rather asymmetric behavior is exhibited by
the  junction. For low magnetic fields an asymmetric
tuning of the flux-flow branch is observed, for moderated magnetic field
ordinary Fiske modes steps are mixed with nonlinear cavity modes steps, and
for quite large magnetic fields a single step  similar to
the velocity-matching step known for uniform width geometries is recovered.

\section*{Acknowledgments}

We acknowledge Professor M. Cirillo for providing us with the
photolithographic masks we used to fabricate the devices.
Fruitful discussions with S. Pagano and C. Nappi, as well as the  financial
support of MURST COFIN00 are also acknowledged.

\end{document}